\documentclass[twocolumn,article,10pt,paper]{IEEEtran}
\usepackage{epsfig,amsfonts,subfigure}
\usepackage[nolist]{acronym}
\usepackage{graphicx,cite,amssymb,amsmath}
\usepackage{color}
\usepackage{psfrag}

\IEEEoverridecommandlockouts

\begin{document}
\begin{acronym}
\acro{LT}{Luby-transform} \acro{SIC}{Successive Interference
Cancellation} \acro{MUD}{Multi User Detection}
\acro{EM}{Expectation-Maximization} \acro{CE}{Channel Estimation}
\acro{SISO}{Soft-Input Soft-Output} \acro{MMSE}{Minimum Mean Square
Error} \acro{CSI}{Channel State Information}
\acro{LLR}{Log-Likelihood Ratio} \acro{APP}{A Posteriori
Probability}
\end{acronym}



\title{On the Derivation of Optimal\\Partial Successive Interference Cancellation}
\author{Francisco L\'azaro Blasco, Francesco Rossetto\\
Institute of Communications and Navigation\\
DLR (German Aerospace Center), Wessling, Germany 82234\\
Email: {\tt Francisco.LazaroBlasco@dlr.de, Francesco.Rossetto@dlr.de}
\vspace{-18pt}}

\maketitle
\date{\today}
\thispagestyle{empty} \setcounter{page}{0}


\begin{abstract}
\textcolor{black}{The necessity of accurate channel estimation for
Successive and Parallel Interference Cancellation is well known.
Iterative channel estimation and channel decoding (for instance by
means of the Expectation-Maximization algorithm) is particularly
important for these multiuser detection schemes in the presence of
time varying channels, where a high density of pilots is necessary
to track the channel. This paper designs a method to analytically
derive a weighting factor $\alpha$, necessary to improve the
efficiency of interference cancellation in the presence of poor
channel estimates. Moreover, this weighting factor effectively
mitigates the presence of incorrect decisions at the output of the
channel decoder. The analysis provides insight into
the properties of such interference cancellation scheme and the
proposed approach significantly increases the effectiveness of
Successive Interference Cancellation under the presence of channel
estimation errors, which leads to gains of up to 3 dB.}

\end{abstract}
\vspace{-1pt}

{\pagestyle{plain} \pagenumbering{arabic}}


\section{Introduction}\label{sec:Intro}

\ac{MUD} has made very significant steps from conceptual tool into
practice~\cite{SIC_Verdu, Hou06, Andrews05, JSAC_MUD_08}. Iterative
\ac{CE} and Interference Cancellation (IC)~\cite{Wang03,
Valenti01:iterativeCE} (for instance by means of the Expectation-Maximization algorithm~\cite{Kobayashi01}) is one of the most appealing techniques,
partly due to its linear complexity in the number of users.
According to this principle, every time new estimates about the
transmitted symbols are available, the channel is estimated again so
as to refine the precision of the Channel State Information (CSI).
On the other hand, as new (hopefully more accurate) CSI is
available, the MUD is repeated, leveraging the CSI improved quality.
Such process is iterated until convergence is achieved. In addition,
the performance of IC is enhanced by the usage of soft-estimates of
the coded symbols based on the computed symbol A Posteriori
Probabilities (\acsp{APP}). If the bits are not
correctly decoded, the estimated bits are regarded
as unreliable and hence little or no signal will be subtracted.

This idea has proved to be very effective for slowly varying
channels, because reliable channel estimation is possible
\cite{Kobayashi01, Wang03, Wehinger06}. The usage of data symbols
for the channel estimation process is in principle even more
attractive for time varying channels, because this approach increases
the sampling frequency of the channel and enables better tracking
thereof. Such type of channels shows up for example with mobile
users or when phase noise is non-negligible. The impact of phase
noise is particularly relevant for instance for applications at high
carrier frequency (say Ku or Ka band, typical in modern satellite
environments), where high stability oscillators can be very
expensive; thus cheap, consumer-grade terminals are significantly
affected by such problem. A pilot based CE would require in these
conditions a too high density of known symbols to accurately track the
channel. The problem of reliable channel estimation is even more
pressing with MUD based on Successive- or Parallel- Interference
Cancellation, since channel estimation errors lead to residual,
non-cancelled interference. While a large number of pilots would
entail an excessive overhead, the amount of symbols used for channel
estimation purposes may be increased by using the output of the
decoding process and therefore iterative CE-MUD approaches sound
very suitable for this task.

On the other hand, a subtle issue with iterative channel estimation
MUD seeps in. The goal of channel decoding is to find the sequence
of coded bits that best fits the received samples, given a set of
known and fixed channel estimates. If the process of channel
estimation and data detection becomes iterative, the receiver will
find the sequence of symbols and channel estimates that
\emph{jointly} fit the received samples. Therefore, the number of
degrees of freedom to {``}explain the observations"
increases and hence the decoded bits will have large \acsp{APP} and will be regarded as reliable even when they do not correspond to the actually transmitted bits. The effectiveness of soft interference cancellation will be severely limited by this problem.

\begin{figure*}[t]
\centering
\psfrag{a0s}{$\alpha_{0,s}$}
\psfrag{a1s}{$\alpha_{1,s}$}
\psfrag{a2s}{$\alpha_{2,s}$}
\psfrag{aus}{$\alpha_{u-1,s}$}
\psfrag{a2s-1}{$\alpha_{2,s-1}$}
\psfrag{aus-1}{$\alpha_{u,s-1}$}
\psfrag{au+1s-1}{$\alpha_{u+1,s-1}$}
\psfrag{aUs-1}{$\alpha_{U-1,s-1}$}
\includegraphics[width=1.8\columnwidth]{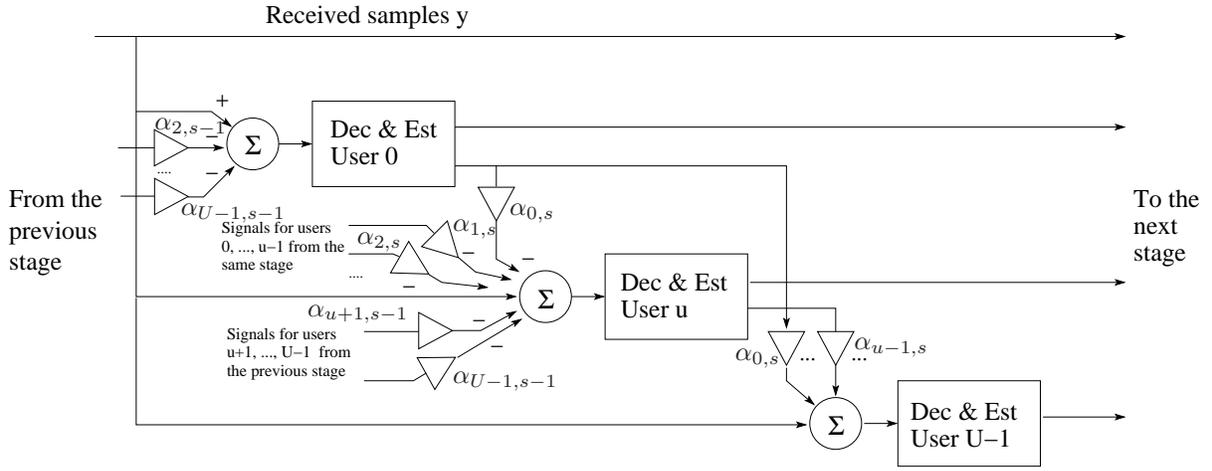}
\caption{The architecture of a generic stage of the multistage SIC receiver.}
\label{fig:receiver_architecture}
\vspace{-10pt}
\end{figure*}

Section~\ref{sec:sys_model} will present the system model, while the
main contribution of this paper is further described in
Section~\ref{sec:alpha} and consists in the definition of a modified
soft interference cancellation scheme that is especially effective
for time-varying channels. The key idea is to multiply by a
coefficient $\alpha$ the reconstructed waveform to be subtracted.
\ac{SIC} is in fact an iterative algorithm to find the solution of a
linear equation system. Therefore, it is equivalent to the
Gauss-Seidel iteration. Applying a weighting factor is equivalent to
Gauss-Seidel over relaxation, and it is known to improve convergence
\cite{Hageman01}. The idea of applying a weighting factor is not new
\emph{per se}, as it has appeared in~\cite{Nambiar99, Divsalar98}.
However, the derivation of the value for this IC factor $\alpha$ was
heuristic and driven by numerical simulation~\cite{Divsalar98}.
Instead, an analytic derivation of such factor is provided and hence
it is possible to optimize its effectiveness and get insight into
the properties of this IC technique. Moreover, our proposed approach
works also for asynchronous users and time varying channels (which
is not the case of~\cite{Kobayashi01}). Section~\ref{sec:simulation}
will evaluate the performance of this scheme by means of simulations
and finally Section~\ref{sec:conc} will draw the conclusions.

\section{System Model}\label{sec:sys_model}

The first part of this section will introduce the essential paper
notation. Then, the basic model that will be further analyzed in
Sections~\ref{sec:alpha} and~\ref{sec:simulation} is described. The
final part (Section~\ref{subsec:Extensions}) will outline some
important extensions to the basic system.

\subsection{Essential Notation}\label{subsec:Common_Symbols}

Let us consider a generic complex number $a$. Its real part,
magnitude, argument and conjugate are denoted as $Re(a)$,
$|a|$, $\mbox{Arg}(a)$ and $a^*$, respectively. The estimate of any
quantity $a$ (complex or real) is denoted as $\hat{a}$.
The Hadamard product (i.e., elementwise) between two vectors $a_1$ and $a_2$ is denoted by $a_1\cdot a_2$. Finally, the expectation of a random variable $a$ is denoted by $\mathbb{E}{[a]}$.

\subsection{Basic Model}\label{subsec:Basic_Model}

The system is composed by $U$ users that transmit simultaneously and
in the same frequency to a common receiver, all involved nodes are
equipped with a single antenna and no spreading is used (a possible
extension to multiple antennas will be outlined in the next
subsection). Each user \mbox{$u\in\{0,\ldots,U-1\}$} generates a
sequence $b_u$ of $B_u$ independent, identically distributed
information bits. Each set of information bits is channel encoded
into $C_u$ coded bits by means of a rate $R_u=B_u/C_u$ channel code.
In the simulation section the UMTS Turbo code will be
assumed~\cite{3GPPcoding}, but in fact the proposed algorithm does
not depend on the specific code, as long as a \ac{SISO} decoder is
adopted~\cite{RyanLinBook2009}. The coded bits are interleaved and
then modulated by means of an $M$-PSK modulation into
\mbox{$X_{d,u}=C_u/\log_2(M)$} modulated data symbols, which are
time multiplexed with $X_{p,u}$ pilots symbols for channel estimation
purposes. Let us denote by $X_u$ the whole set of data and pilot symbols. They are pulse modulated
and transmitted through a frequency flat AWGN channel, which adds
complex zero mean circularly symmetric White Gaussian Noise (WGN)
with per dimension variance $\sigma_n^2$. The channel is not assumed
to be slowly time varying, i.e., it may change from symbol to symbol
due to, for example, phase noise or Doppler. We shall assume that
the former is indeed present due to the instability of the
transmitters and receiver oscillators.

The signals from the $U$ different transmitters arrive at the
receiver. The users will be assumed in the description of the system
model to be symbol and frame synchronous for notational simplicity.
This assumption will also make the derivation and discussion of the
$\alpha$ factor easier to follow in Section~\ref{sec:alpha}.
However, all simulations have been carried out for the symbol and
frame asynchronous case, thus the performance evaluation is
realistic in this respect.

The receiver performs matched filtering and sampling and obtains a sequence of samples $y_t$, where $t$ is the discrete temporal index, $0\leq t\leq\max(\{X_u\})-1$:

\begin{equation}
y_t = \sum_{u=0}^{U-1}h_{t,u}x_{t,u} + w_t
\end{equation}
where $h_{t,u}=|h_{t,u}|e^{i\theta_{t,u}}$ is the frequency flat
complex channel at time $t$ for user $u$, $x_{t,u}$ is the modulated
symbols transmitted at time $t$ by user $u$ and $w_t$ is the complex
WGN. \textcolor{black}{ In our case the variability of the channel
coefficient $h_{t,u}$ is induced by the phase noise due to
oscillator instabilities. However, the analysis presented is valid
for any flat complex channel.} A random-walk (Wiener) phase noise
model with Gaussian increments is adopted~\cite{Colavolpe05}. The
additional phase due to such impairment can be described as:

\begin{equation}
\theta^{(p)}_{t,u} = \theta^{(p)}_{t-1,u} + \Delta_{t,u}
\end{equation}

The increment $\Delta_{t,u}$ is a zero mean real Gaussian random
variable with variance $\sigma_p^2$. The receiver performs a
multistage \ac{SIC} with iterative CE and Channel Decoding (CD),
which is depicted in Fig.~\ref{fig:receiver_architecture}. In order
to improve the error correction performance, the users are ordered
in descending order of power~\cite{SIC_Verdu} and we assume without
loss of generality that the user index $u$ corresponds to the
decoding order. At each stage $s, 0\leq s \leq S-1$ the $u$-th user
is decoded based on the following signal: \vspace{-5pt}
\begin{eqnarray}
y_{t,s,u} = y_t &-& \sum_{u1=0}^{u-1}\alpha_{s,u1}\hat{h}_{t,s,u1}\hat{x}_{t,s,u1} -\nonumber \\
                &-& \sum_{u1=u+1}^{U-1}\alpha_{s-1,u1}\hat{h}_{t,s-1,u1}\hat{x}_{t,s-1,u1}
\end{eqnarray}
which is equal to the received samples minus the estimated
interference from the other users. The interference from the other
users is reconstructed using the \ac{APP} rather than extrinsic
probabilities. It has been shown that when a graph based approach is
used for \ac{MUD} extrinsic probabilities should be used rather than
\ac{APP} \cite{boutros02}. However it is not known whether \ac{APP}
or extrinsic probabilities should be used in case a different
\ac{MUD} algorithm is used, such as \ac{SIC} \cite{Rasmussen01}. In
our case \ac{APP} provide better performance. Note that
$y_{t,0,0}=y_t$. The estimates of the channel and of the modulated
symbols from stage $s$ of user $u$ at time $t$ are denoted by
$\hat{h}_{t,s,u}$ and $\hat{x}_{t,s,u}$. Estimates from the same
stage $s$ are employed for users $0, ..., u-1$ already decoded in
this stage, whereas estimates from the previous stage $s-1$ are used
for the other users $u+1, ..., U-1$. Note the presence of the
factors $\alpha_{s,u}$, which will be the focus of the next section.
Their purpose is to weight the confidence on the estimated waveform.

The values for $\hat{h}_{t,s,u}$ and $\hat{x}_{t,s,u}$ are derived
from the channel estimation and channel decoding, respectively. In
order to fight the channel variability, an iterative channel
estimation/channel decoding algorithm which follows the
Expectation-Maximization (EM) approach is adopted~\cite{Kobayashi01,
Dempster77:EM, Boutros05, herzet01:turbo_synch}. The derivation of
the steps is carried out according to~\cite{Boutros05} and it can be
shown that the E-step is simply the channel decoding. From the
E-step, the \acp{APP} of the coded bits $c_{t,u}$ can be
evaluated~\cite{RyanLinBook2009} and hence $\mathbb{E}[x_{t,s,u}]$
\footnote{The expectation is actually conditioned to the
observation, $\mathbb{E}[x_{t,s,u}|y_t]$. In the following the
conditioning on the observation will be dropped for notational
simplicity.} and $\mathbb{E}[|x_{t,s,u}|^2]$ are computed for the
sake of channel estimation and interference cancellation. Moreover,
the channel decoder gives as additional output $\hat{x}_{t,s,u}$,
which can be (but need not be) $\mathbb{E}\left[ x_{t,s,u}\right ]$,
i.e., the \ac{MMSE} estimate of $x_{t,s,u}$. In addition,
$\mathbb{E}\left[|x_{t,s,u}|^2\right]$ will be denoted with a slight
abuse of notation $|\hat{x}_{t,s,u}|^2$. Note that also known
symbols (for instance the pilots) are employed in the CE process.
Since these symbols are known with no uncertainty,
$\mathbb{E}[x_{t,s,u}]=x_{t,u}$ and
$\mathbb{E}[|x_{t,s,u}^2|]=|x_{t,u}^2|$.

On the other hand, the M-step corresponds to the channel estimation carried out according to these equations~\cite{Boutros05}:
\vspace{-5pt}
\begin{eqnarray}
\hat{|h|}_{t,s,u}      &=& \frac{\sum_{t1=t-W}^{t+W}Re\left(\hat{x}_{t1,s,u}^*y_{t1,s,u}\right)}{\sum_{t1=t-W}^{t+W}|\hat{x}_{t1,s,u}^2|} \label{eq:chanMagn}\\
\hat{\phi}_{t,s,u}     &=& \mbox{Arg}\sum_{t1=t-W}^{t+W}\hat{x}_{t1,s,u}^*y_{t1,s,u}                  \label{eq:chanPhase} \\
\hat{h}_{t,s,u}        &=& \hat{|h|}_{t,s,u}e^{i\hat{\phi}_{t,s,u}}                                      \label{eq:chan} \\
\hat{\sigma}_{n;s,u}^2 &=& \frac{1}{2X_u}\sum_{t1=0}^{X_u-1}|y_{t1,s,u}-\hat{h}_{t1,s,u}\hat{x}_{t1,s,u}|^2 \label{eq:noiseAndIntVariance}
\end{eqnarray}
\vspace{-5pt}

Note that the channel is time variant and therefore all estimates
are performed over a sliding window of size $2W+1$ samples around
the desired time index $t$, where $W$ depends on the coherence time
of the channel.

After the $S$ MUD stages, the channel decoding estimates of the
information bits $\hat{b}_u$ are hard thresholded and given as
output of the MUD/CD process.

\vspace{-5pt}
\subsection{Extensions}\label{subsec:Extensions}

We briefly outline here two extensions of this framework. The first
one concerns symbol asynchronous users and the latter
V-BLAST~\cite{Tse08}.

In the first case, the users are symbol asynchronous and the
receiver first samples the overall signal into discrete time samples
$\{y_t\}$ with a timing not necessarily related to the sampling time
of the users. Such timing will be called ``the
common time frame" and is opposed to the optimal matched filter
sampling choice for each user, that will be called the
``optimal single user time frame". The channel
estimation and decoding block receives samples at the common time
frame, which are interpolated to the optimal single user frame. All
operations that are related to the tagged user (e.g., CE and CD) are
performed in the latter system and the computed channel and symbol
estimates are then interpolated back to the common time frame. After
this operation, the reconstructed waveform can be subtracted from
the received samples $y_t$.

The second extension concerns V-BLAST. The receiver and the
transmitters are assumed to be equipped with $N_r$ and $M_t$
antennas, respectively, and the transmitters send independent
streams. The users are ordered according to some usual criterion
(say, maximum post-processing SNR) and the received samples are
first processed by means of Zero Forcing or \ac{MMSE} filtering and
then the selected user is decoded by means of the previously
described iterative channel estimation and decoding approach. Once
the estimates are ready, the weight factor $\alpha_{s,u}$ is
computed, the waveform of the detected user is reconstructed,
multiplied by $\alpha_{s,u}$ and subtracted from the received
samples. This approach is reminiscent of Turbo MIMO~\cite{Haykin04},
with the difference that also channel estimation is iterative and
for the presence of the $\alpha$ factors.

\section{Optimal Partial Interference Cancellation} \label{sec:alpha}

Let us denote as $y_u=h_u \cdot x_u$ the noiseless signal received
from user $u$,  being $x_u$ and $h_u$ the symbols and channel
coefficients from user $u$. Stage and temporal indexes are dropped
here for notational simplicity.  Let $\hat{y}_u = \hat{h}_u \cdot
\hat{x}_u$ be an estimate of $y_u$. Assume now that we want to
cancel the interference caused by user $u$ on other users. In
accordance to the model defined in Section~\ref{sec:sys_model} a
\ac{SIC} scheme is applied in which the estimates are first
multiplied by a weighting factor $\alpha_u$ and then subtracted,
with $0 \leq \alpha_u \leq 1$. The interference cancellation
efficiency~\cite{sambhwani01:uplinkIC} for such a scheme can be
defined as:
\begin{equation}
 \beta = 1 - \mathbb{E} \left[ \frac{|y_u -  \alpha_u \hat{y}_u|^2}{ {|y_u|}^2} \right]
\end{equation}

In an ideal case, when the interference caused by $y_u$ is
completely removed, $\beta = 1$. On the other hand, if no
interference is removed, $\beta = 0$.

If the adopted modulation is PSK, the interference cancellation efficiency can be expanded as:
\begin{eqnarray}
\beta &=& 1 - \mathbb{E} \left[ \frac{|h_u \cdot x_u -  \alpha_u \hat{h}_u \cdot \hat{x}_u|^2}{ {|h_u\cdot x_u|}^2} \right] \nonumber \\
      &=& 1 - \mathbb{E} \left[ \frac{|h_u \cdot x_u  - \alpha_u  h_u \cdot \hat{x}_u + \alpha_u  h_u \cdot \hat{x}_u -  \alpha_u \hat{h}_u \cdot \hat{x}_u|^2}{ {|h_u|^2\cdot |x_u|}^2} \right] \nonumber \\
      &=& 1 - \mathbb{E} \left[ |x_u - \alpha_u \hat{x}_u|^2 \right] - \alpha_u^2 \mathbb{E} \left[ |\hat{x}_u|^2 \right]
      \mathbb{E} \left[ \frac{|h_u-\hat{h}_u|^2}{|h_u|^2} \right] \nonumber \\
&& -\mathbb{E} \left[ \frac{ 2 \alpha_u Re \{ h_u^* \cdot (x_u-\alpha_u \hat{x}_u)^* \cdot (h_u - \hat{h}_u) \cdot \hat{x}_u\} } {|h_u|^2} \right] \label{eq:alpha_1}\\
      &\simeq& 1 - \mathbb{E} \left[ |x_u - \alpha_u \hat{x}_u|^2 \right] - \alpha_u^2 \mathbb{E} \left[ |\hat{x}_u|^2 \right]
      \mathbb{E} \left[ \frac{|h_u-\hat{h}_u|^2}{|h_u|^2} \right] \label{eq:alpha_2}
\end{eqnarray}

The assumption of a PSK modulation implies that $|x_u|^2=1,\forall
x_u$. Extension of such analysis for non-constant envelope
modulations is left for future work.

The last step from Eq.~(\ref{eq:alpha_1}) to
Eq.~(\ref{eq:alpha_2}) is justified because the last term of
Eq.~(\ref{eq:alpha_1}) is generally zero under the reasonable
assumption that the channel estimation error is uncorrelated to the
value of the channel. Under these conditions:
\begin{eqnarray}
\beta&=& 1 - ( 1 + \alpha_u^2 \mathbb{E} \left[ |\hat{x}_u|^2 \right] - 2 \alpha_u \mathbb{E} \left[ Re\{ x_u \cdot \hat{x}_u^*\} \right] ) \nonumber\\
     &-& \alpha_u^2 \mathbb{E} \left[ |\hat{x}_u|^2 \right] \bar{E}
\end{eqnarray}
where $\bar{E} = \mathbb{E} \left[ \frac{|h_u-\hat{h}_u|^2}{|h_u|^2}
\right]$ is the normalized mean square channel estimation error. Note that the second term in Eq.~\ref{eq:alpha_2} is related to the reliability of the channel decoding, while the third term conveys information on the channel estimation accuracy.

Setting the first derivative of $\beta$ with respect to $\alpha_u$
to $0$, it can be easily verified that the value $\alpha_{u{,opt}}$
that maximizes $\beta$ is:
\begin{equation}
 \alpha_{u,opt} = \frac{\mathbb {E} \left[ Re\{ x_u \cdot \hat{x}_u^* \} \right] }{ \mathbb{E} \left[ |\hat{x}_u|^2 \right] }\frac{1}{1 + \bar{E}}
\label{eq:opt_alpha}
\end{equation}

Note that the optimal value of the weighting coefficient
$\alpha_{u,opt}$ depends on $\mathbb{E} \left[ |\hat{x}_u|^2
\right]$, $\bar{E}$ and $\mathbb {E} \left[ Re\{ x_u \cdot
\hat{x}_u^* \} \right]$. The first term can be computed at the
receiver, since $\hat{x}_u$ is available at the output of the
\ac{SISO} decoder. It can be assumed that the second term can be
estimated given the SINR at which the receiver operates. The
computation of $\mathbb {E} \left[ Re\{ x_u \cdot \hat{x}_u^* \}
\right]$ involves perfect knowledge of the transmitted symbols.
However, a good estimate of this term can be obtained replacing
$x_u$ by hard decisions on $\hat{x}_u$. \textcolor{black}{$\bar{E}$
can be also reliably estimated by means of a look-up table based on
an estimate of the SINR.}

Furthermore, note that the right hand side of
Eq.~(\ref{eq:opt_alpha}) can be described by two terms, the first
depending on the reliability of the channel decoder output, while
the second depends on the accuracy of the channel estimation.  When
pilot aided \ac{CE} is considered, the optimal weighting
coefficients for data and pilot symbols are different. In case of
data symbols $\alpha_{opt}$ is given by:
\begin{equation}
 \alpha_{u_{opt}}^{(d)} = \frac{\mathbb {E} \left[ Re\{ x_{u}^{(d)} \cdot \hat{x}_{u}^{(d)^*} \} \right] }{ \mathbb{E} \left[ |\hat{x}_{u}^{(d)}|^2 \right] ( 1 + \bar{E}) }
\label{eq:opt_alpha_data}
\end{equation}
where $x_{u}^{(d)}$ are the data symbols of user $u$. In case of the
pilot symbols given that their value is known a priori, Eq.
\eqref{eq:opt_alpha_data} can be simplified to:
\begin{equation}
 \alpha_{u_{opt}}^{(p)} = \frac{1}{ 1 + \bar{E} }
 \label{eq:opt_alpha_pilot}
\end{equation}

Simulations verified that $ \alpha_{u_{opt}}^{(p)} \geq
\alpha_{u_{opt}}^{(d)}$, with equality for high SNR. For low SNR,
the value of the data symbols can not be reliably estimated , $x_u$
and $ \hat{x}_u$ are uncorrelated and therefore: $\mathbb {E} \left[
Re\{ x_{u}^{(d)} \cdot \hat{x}_{u}^{d^*} \} \right] \simeq 0$. On
the other hand, when the estimates become more reliable, $\mathbb
{E} \left[ Re\{ x_{u}^{(d)} \cdot \hat{x}_{u}^{d^*} \} \right]
\simeq \mathbb{E} \left[ |\hat{x}_{u}^{(d)}|^2 \right]$, and
therefore $ \alpha_{u_{opt}}^{(p)} \simeq \alpha_{u_{opt}}^{(d)}$.

Let us now consider the case in which $ \hat{x}_u = x_u$. This is
the case of pilot symbols, or data symbols when we operate in a
region where the Bit Error Rate (BER) $\simeq 0$. Let us now assume that the normalized
channel estimation error, $\bar{E}$, is known. We shall call full
\ac{SIC} the conventional \ac{SIC} scheme with $\alpha = 1$. Under
full \ac{SIC} the normalized interference caused by user $u$ to
other users can be expressed as:
\begin{equation}
 \bar { \mathcal{I} }_{full} =\mathbb {E} \left[ \frac{|y_u -  \hat{y}_u|^2}{ {|y_u|}^2} \right]= \mathbb {E} \left[ \frac{|h_u \cdot x_u -  \hat{h}_u \cdot x_u|^2}{ {|h_u \cdot x_u|}^2} \right] = \bar {E}
 \label{eq:I_full}
\end{equation}

In a system with optimal partial \ac{SIC}, according to Eq.
\eqref{eq:opt_alpha}, the normalized interference can be expressed
as:
\begin{equation}
 \bar{ \mathcal{I} }_{partial} = \mathbb {E} \left[ \frac{|y_u -  \alpha_{u_{opt}}\hat{y}_u|^2}{ {|y_u|}^2} \right]= \frac {\bar {E}}{1+\bar{E}}
 \label{eq:I_partial}
\end{equation}

We can define $ \bar{ \gamma } $ as the ratio of the normalized
interference with full \ac{SIC} and optimal partial \ac{SIC}:
\begin{equation}
\bar{ \gamma } = \frac{ \bar{ \mathcal{I}}_{partial}} {\bar{ \mathcal{I}}_{full}}= \frac { 1 } { 1+\bar{E} }
 \label{eq:I_diff}
\end{equation}

Note that since $\bar{E} \geq 0$, $ \bar{ \gamma } $ is a
monotonically decreasing function of $\bar{E}$ and $ \bar{ \gamma }
\leq 1$.

Let us now consider the opposite case, where the channel estimates
are perfect, but the estimates of the data symbols are not. Let
$\lambda_{u}$ be the \acp{LLR} of the \acsp{APP}
obtained at the output of the \ac{SISO} channel decoder of user $u$.
For BPSK, the symbol estimates are calculated as:
\begin{equation}
\hat{x}_u= \tanh \left( \frac{\lambda_u}{2} \right)
 \label{eq:soft_estimates}
\end{equation}

\begin{figure}[t]
\centering
\includegraphics[width=0.82\columnwidth]{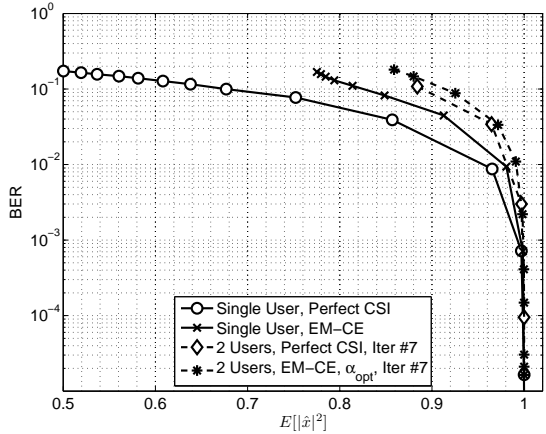}
\vspace{-4pt}
\caption{BER vs. soft values in a 1 user system (solid lines) and 2 user system (dashed lines) with 2 dB power imbalance.}
\label{fig:BER_vs_soft}
\vspace{-10pt}
\end{figure}

This choice is optimal from a \ac{MMSE} consideration \cite{Wang03}.
In other words, under perfect \ac{CSI} and using soft decision
according to Eq. \eqref{eq:soft_estimates}, full \ac{SIC}
$(\alpha=1)$ minimizes $\beta$. This was verified by means of
simulations. However, when \ac{EM} based \ac{CE} is used, this is no
longer true. Fig. \ref{fig:BER_vs_soft} shows the evolution of the
BER vs. $\mathbb{E} \left[ |\hat{x}_{u}^{(d)}|^2 \right]$ for
perfect \ac{CSI} and EM based \ac{CE}
 using the UMTS turbo code with code rate $0.53$. The figure shows
 results for a single user case and a two user case according to the
 setup described in Section~\ref{sec:simulation} (user $2$). For the single
 user case it can be seen how for the same
BER, $\mathbb{E} \left[ |\hat{x}_{u}^{(d)}|^2 \right]$ is higher for EM
based \ac{CE} than for perfect \ac{CSI}. This means that the system
is overconfident in the decisions it takes. The figure also shows
how in the presence of two users $\mathbb{E} \left[
|\hat{x}_{u}^{(d)}|^2 \right]$ is higher than in the single user case,
both with perfect \ac{CSI}. Finally the two user case with EM based
\ac{CE} produces the highest $\mathbb{E} \left[ |\hat{x}_{u}^{(d)}|^2
\right]$.

In Section \ref{sec:simulation} it will be shown how optimal partial
\ac{SIC} can translate into a considerable performance improvement
in a system where \ac{SIC} with \ac{EM} \ac{CE} is used. Especially
when the power imbalance between users is low and hence the channel
estimation error is high, full \ac{SIC} is unable to bootstrap due
to the high residual interference from other users. Optimal partial
\ac{SIC} decreases the residual interference enabling gains of up to
$3$ dB for power imbalance $1$ dB.

\section{Simulation Study}\label{sec:simulation}
A simulation study was carried out in order to evaluate the
performance of the proposed partial interference cancellation
scheme. A simulation setting with two symbol and frame asynchronous
users is used. The channel code is the UMTS turbo code
\cite{3GPPcoding}. The code rates of user $1$ and user $2$ are $0.72$
and $0.53$ respectively.\footnote{Such rates corresponds to the
capacity of a BPSK constrained AWGN channel with nominal SNR for the
second user 1.5 dB and power imbalance 2 dB. The SIR of the second
user is computed as 2 dB, that is to say, IC of the first user leads
to a 5 dB improvement. Finally, the channel codes are assumed to be
1 dB away from capacity.} The number of \ac{SIC} stages is set to
$7$ in all cases, as no meaningful performance improvement is
observed after this point. In each stage each user performs $15$
iterations of the \ac{EM} algorithm. Both transmitters employ the
same random interleaver. The modulation used is BPSK for both users,
and the code block length is $C_u=5000$ symbols, $\forall u$. The
number of pilot symbols per code block is $X_{u,p}=256,\forall u$.
Pilot symbols are boosted by $3$ dB with respect to data symbols.
The channel is assumed to be AWGN and the phase noise increase
between two consecutive symbols has a variance $\sigma_p^2= 100/f_0$
rad$^2$/symbol, where $f_0$ is the baud rate (in our
simulations 10 kBaud).\footnote{This statistics has been derived from
satellite transceivers normally available to the authors.}
Different simulations were carried out for power imbalance,
$P_{U,u_1,u_2}$, ranging from $1$ dB to $7$ dB, being user $1$
always the strongest one. In contrast to much of the work on
\ac{MUD}, no spreading is employed.\footnote{ In spite of the use of
a low cardinality constellation like BPSK and the fact that no
spreading is used, the interference caused among users can be
considered Gaussian due to
 phase noise and the effect of the interleaver.}

\begin{figure}[t]
\centering
\includegraphics[width=0.82\columnwidth]{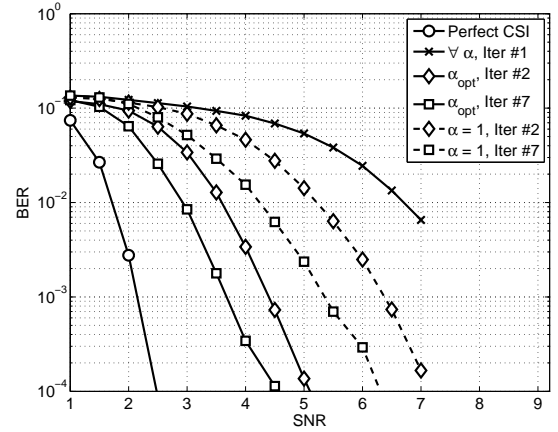}
\vspace{-4pt}
\caption{BER vs. SNR for the first user, power imbalance $2$ dB.}
\label{fig:BER_vs_SNR_user_1}
\vspace{-5pt}
\end{figure}

\begin{figure}[t]
\centering
\includegraphics[width=0.82\columnwidth]{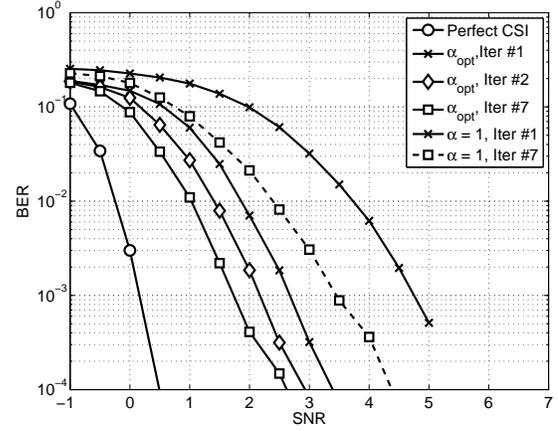}
\vspace{-4pt}
\caption{BER vs. SNR for the second user, power imbalance $2$ dB.}
\label{fig:BER_vs_SNR_user_2}
\vspace{-10pt}
\end{figure}

Our scheme is compared to multistage full \ac{SIC} with EM based CE (very akin to~\cite{Kobayashi01}). Fig. \ref{fig:BER_vs_SNR_user_1} and Fig.
\ref{fig:BER_vs_SNR_user_2} show the BER of user $1$ and $2$,
respectively, for a power imbalance of $2$ dB. If we analyze the SNR
at which the different schemes reach a BER of $1e-4$, for both users
optimal partial \ac{SIC} results in a SNR gain of
$2$ dB with respect to full \ac{SIC}. Moreover, we can see how
optimal partial \ac{SIC} with $2$ stages outperforms full \ac{SIC}
with $7$ stages. However, both \ac{SIC} schemes with estimated
\ac{CSI} are far from the performance achieved by perfect \ac{CSI}.
For optimal partial \ac{CSI} the loss with respect to perfect CSI is
$2$ dB for both users, half of the gap of the state-of-the art system.

\begin{figure}[t]
\centering
\includegraphics[width=0.82\columnwidth]{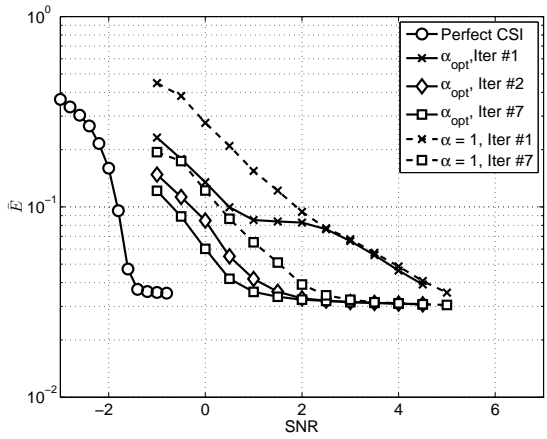}
\vspace{-4pt}
\caption{Normalized mean square error in the channel estimate $\bar{E}$ for the second user with a 2 dB power imbalance between the two users.}
\label{fig:CEmse_vs_SNR}
\end{figure}

Fig. \ref{fig:CEmse_vs_SNR} shows the evolution of $\bar{E}$ with
the SNR for user $2$. It can be seen how after $7$ \ac{SIC} stages
the scheme with optimal partial \ac{SIC} outperforms full \ac{SIC}.
However the performance of both schemes is far from the performance
of the single user case. The reason for this is the remaining
interference from user $1$. It is also quite remarkable that optimal partial \ac{SIC} after $2$ stages outperforms
again full \ac{SIC} with $7$ stages. Another important fact is that
there is a floor at $\bar{E} = 0.031$, which is caused by the
presence of strong phase noise. Such problem clearly limits the
performance of \ac{SIC} since the interference can never be
completely canceled. The evolution of $\bar{E}$ for user $1$, which
is not shown in any figure, is very similar to that of user $2$,
presenting also the same floor.

\begin{figure}[t]
\centering
\includegraphics[width=0.82\columnwidth]{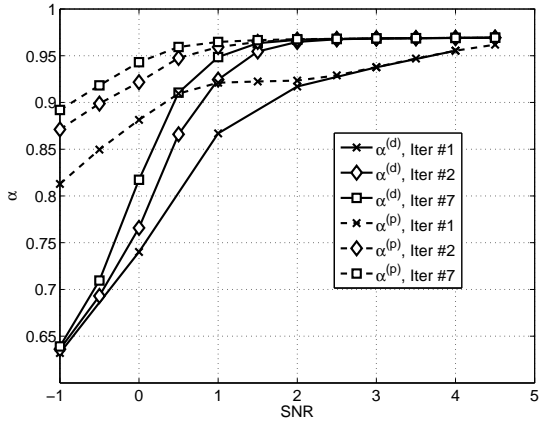}
\vspace{-4pt}
\caption{Evolution of the $\alpha$ factor for the second user as a function of the SNR. The power imbalance is 2 dB.}
\label{fig:alpha_vs_SNR}
\vspace{-10pt}
\end{figure}

The evolution of $\alpha_u^{(d)}$ with respect to the SNR is shown in
Fig.~\ref{fig:alpha_vs_SNR}. It can be seen how $\alpha_u^{(d)}$
increases with the number of \ac{SIC} stages. The figure also shows
how the maximum value of $\alpha_u^{(d)}$ is $0.97<1$. The maximum possible value (i.e., 1) is not reacheable because of
the presence of a floor in the channel estimation. The proposed scheme is aware of this floor and avoids full removal of the estimated signal.

\begin{figure}[t]
\centering
\includegraphics[width=0.82\columnwidth]{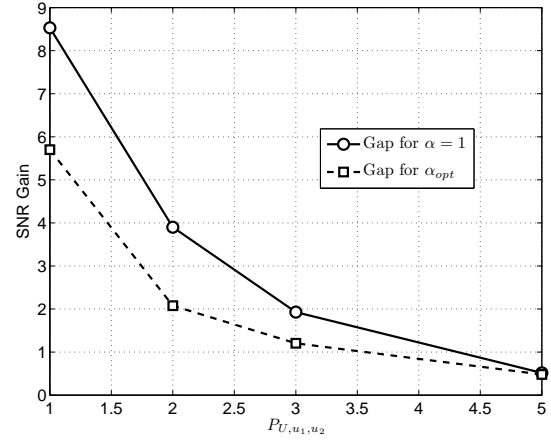}
\vspace{-4pt}
\caption{Additional SNR required by the MUD schemes with respect to perfect CSI as a function of the power imbalance.}
\label{fig:gain_vs_PU}
\vspace{-10pt}
\end{figure}

Finally, Fig.~\ref{fig:gain_vs_PU} shows the SNR at which optimal partial \ac{SIC}
and full \ac{SIC} reach BER $1e-4$ for different power imbalances.
It can be seen how the gain provided by optimal partial \ac{SIC} is
high for low power imbalance and decreases as the power imbalance
increases. Note that in many cases the optimal choice of $\alpha$ halves the gap to perfect CSI given this time varying channel.

\section{Conclusions}\label{sec:conc}
The principle of soft \ac{SIC} hinges on the idea that a low confidence on the decoded bits would reduce the amount of subtracted interference. The EM algorithm is known to be an effective tool to improve the performance of MUD. However, it has the side effect of leading to overconfidence in the \acp{LLR} which has a negative effect on the performance of \ac{SIC}.
The optimal weighting factor $\alpha$ for \ac{SIC} has been derived, which is able to partially compensate for this overconfidence in the \acp{LLR}. Moreover, it is also able to mitigate the negative impact of imperfect \ac{CSI}. The proposed technique shows gains up to 3 dB when the detected users are received with comparable powers. Furthermore, the performance gap between perfect CSI and the state of the art is roughly halved by the suggested scheme.

\section{Acknowledgments}
The authors would like to acknowledge Dr. G. Liva and G. Garramone
for the useful discussions.

The authors are supported inpart by the Space Agency of the German Aerospace Center and the Federal Ministry of Economics and Technology based on the agreement of the German Bundestag with the support code 50 YB 0905 and in part by the ESA Contract No. 23030/10/NL/CLP "NICOLE".

\bibliographystyle{IEEEtran}
\bibliography{IEEEabrv,studio3}


\end{document}